# Upper Rhine Valley: A migration crossroads of middle European oaks

**Authors:** Charalambos Neophytou & Hans-Gerhard Michiels

**Authors' affiliation:**

Forest Research Institute (FVA) Baden-Württemberg
Wonnhaldestr. 4
79100 Freiburg
Germany

**Author for correspondence:** Charalambos Neophytou

**Postal address:**
Forest Research Institute (FVA) Baden-Württemberg
Wonnhaldestr. 4
79100 Freiburg
Germany

**Telephone number:** +49 761 4018184

**Fax number:** +49 761 4018333

**E-mail address:** chneophytou@gmail.com

**Short running head:** Upper Rhine oak phylogeography

**ABSTRACT**

The indigenous oak species (*Quercus* spp.) of the Upper Rhine Valley have migrated to their current distribution range in the area after the transition to the Holocene interglacial. Since post-glacial recolonization, they have been subjected to ecological changes and human impact. By using chloroplast microsatellite markers (cpSSRs), we provide detailed phylogeographic information and we address the contribution of natural and human-related factors to the current pattern of chloroplast DNA (cpDNA) variation. 626 individual trees from 86 oak stands including all three indigenous oak species of the region were sampled. In order to verify the refugial origin, reference samples from refugial areas and DNA samples from previous studies with known cpDNA haplotypes (chlorotypes) were used. Chlorotypes belonging to three different maternal lineages, corresponding to the three main glacial refugia, were found in the area. These were spatially structured and highly introgressed among species, reflecting past hybridization which involved all three indigenous oak species. Site condition heterogeneity was found among groups of populations which differed in terms of cpDNA variation. This suggests that different biogeographic subregions within the Upper Rhine Valley were colonized during separate post-glacial migration waves. Genetic variation was higher in *Quercus robur* than in *Quercus petraea*, which is probably due to more efficient seed dispersal and the more pronounced pioneer character of the former species. Finally, stands of *Q. robur* established in the last 70 years were significantly more diverse, which can be explained by the improved transportation ability of seeds and seedlings for artificial regeneration of stands during this period.

## 1. INTRODUCTION

The Upper Rhine Valley is one of the major geographic features in the western part of Central Europe. As a biogeographical region, it displays a relatively wide variety of ecological conditions, hosting diverse vegetation communities, which include several types of broadleaved forests (Oberdorfer, 1992). Due to favourable climate and topography, human settlement in the area began early in the Holocene and land use dates back to at least 7500 years before present (BP; Lang et al., 2003; Zolitschka et al., 2003; Houben et al., 2006). Oaks (*Quercus* spp.) have been both an important member of the natural vegetation communities and a source of timber and other non-wood products for humans during much of the Holocene (since 10,500 years BP). The genus *Quercus* is represented by three species of white oaks (section Quercus based on recent taxonomic studies; Denk and Grimm, 2010) in the area. *Quercus robur*, being able to tolerate waterlogging, mainly occupies lowland areas flanking the river banks (Aas, 2008a). On the other hand, *Q. petraea* requires well drained soils and thus is more abundant in hilly areas like the piedmont of the Vosges Mountains (Aas, 2008b). The third species, *Quercus pubescens*, is a submediterranean vegetation element. In the Upper Rhine Valley, it has a marginal distribution occurring in exceptionally dry sites with alkalic soils (Oberdorfer, 1992; Bussotti, 2008).

As many arboreal elements of the regional flora, oaks migrated to the Upper Rhine Valley soon after the end of the last ice age, during the Preboreal period (first occurrence about 9500 to 9000 years BP), originating from refugial areas around the Mediterranean Sea (Brewer et al., 2002). Genetic evidence about the refugial origin and post-ice-age recolonization routes of Central European oaks has been provided by a large number of studies conducted in the last two decades, mostly based on chloroplast DNA (e.g. Petit et al., 1993, Petit et al., 2002a; Deguilloux et al., 2004). In general, it has been shown that oak populations in Central Europe share their chloroplast DNA haplotypes (chlorotypes) with regions around the Mediterranean, which hosted refugial populations during the last glacial. The genetic differentiation among refugial sites agrees with the hypothesis that the main maternal lineages of cpDNA arose there during the last glacial (or even earlier) under the effects of genetic drift and mutation (Kremer and Petit, 1993; Petit et al., 2002a). The longitudinal distribution of the different maternal lineages reflects a general south-to-north direction of post-glacial recolonization pathways (Petit et al., 2002a, Slade et al., 2008). Moreover, the general lack of further genetic variants confined to non-refugial regions supports that no significant genetic differentiation has happened during the current interglacial (Petit et al., 2002a).

Post-ice-age migration history has largely determined regional cpDNA diversity levels of the Central European oaks. High cpDNA diversity arose where different recolonization routes met (Csaikl et al., 2002; Petit et al., 2002b). On the other hand, seed mediated gene flow, being restricted by the barochoric seed dispersal of oak acorns, contributed to a limited within-population differentiation, but strong phylogeographic structure in terms of cpDNA (Ennos, 1994, Petit et al., 2004). An additional major factor affecting cpDNA variation in oaks is hybridization. Due to linkage equilibrium, chlorotypes can be exchanged among interfertile species after an initial hybridization event and a number of successive backcrossings, (Kremer and Petit, 1993). A wide scale sharing of cpDNA geographic patterns supports historical interbreeding among members of the European white oaks (Petit et al., 2002b). Finally, human management may have also had important consequences on cpDNA variation and spatial distribution. Through seed transfer and creation of artificial stands, human has intervened into seed mediated cpDNA gene flow, by increasing it and probably by contributing to a decrease of differentiation among populations (König et al., 2002; Gailing et al., 2007). This human impact is expected to be strong in areas with an early human settlement and intensive land use, like the Upper Rhine Valley.

In the present study, we focus on the phylogeography of the indigenous oaks in the Upper Rhine Valley based on chloroplast DNA markers. First, we investigate the refugial origin of the stands and the spatial distribution of the different genetic variants in a region with complex migration history. Second, we analyze the genetic variation and its distribution within and among populations and we compare its patterns between *Q. robur* and *Q. petraea*, taking their ecological and physiological differences into account. Third, we address the human influence on the cpDNA distribution of the oaks, by putting a special emphasis on the consequences of recent forest management.

## 2. MATERIALS AND METHODS

### 2.1. Study area and sample collections

The study area comprises the Upper Rhine Valley, delimited by the Vosges Mountains to the west, the Jura Mountains to the south and the Black Forest to the east. The northern limit of the study area was put along the border between Alsace (France) and Rhineland-Palatinate (Germany), along the Rhine (separating Rhineland-Palatinate from Baden-Württemberg) and between Baden-Württemberg and Hessen. The sampled forest stands mostly occurred in lowland sites and to a lesser extent in the foothill zone. A total of 626 trees from 86 oak stands were sampled. The sample was representative of the whole geographic region and covered the whole amplitude of site conditions. Individuals of all three indigenous oak species of the area – *Quercus robur*, *Quercus petraea* and *Quercus pubescens* – were included. Classification into species groups or intermediates was made based on morphological (field characterization) and genotypic data (Bayesian clustering analyses based on 21 nuclear microsatellite loci; results not shown here). In 13 cases, individuals of more than one species were sampled from the same stand. Due to the fact that *Quercus robur* is more widespread in the area, it was represented by a higher number of individuals and stands. In total, 314 individuals of *Q. robur* were collected from 63 stands compared to 214 *Quercus petraea* individuals from 39 stands and 90 *Quercus pubescens* individuals from six stands. In addition, sampling included eight hybrid forms found in seven different stands. Leaf material or buds were collected from each tree. Information about the age of the stands was gathered based on forest local inventory data.

With respect to their site and climatic conditions, the sampled stands were subdivided into three biogeographical subregions (Table 1). A further categorization was made concerning the age of the stands. Again three categories were defined: (1) Stands of age up to 70 years, which have originated after the Second World War, (2) stands of age from 100 to 130 years, due to the fact that seed transportation by rail in Central Europe reached a maximum in this period (König et al., 2002, Gailing et al., 2007), and (3) stands older than 130 years. Stands with an age between 70 and 100 years were not represented in our sample, probably due to reduced plantation activity between the First and the Second World War. Details about the sample stands are included in the supplementary material (supplementary data in electronic version; Table S1).

In addition to the sampled stands in the Upper Rhine Valley, 71 trees from 11 oak stands in the Iberian Peninsula (Navarra and Basque Country), Italy (Sardinia and Central Italy) and the Balkan Peninsula (Central Bulgaria and Northern Greece), representing refugial sites, were included in the sample. Finally, 7 DNA reference samples representing the most common central European haplotypes of white oaks (section Quercus according to a recent study of Denk and Grimm, 2010), kindly provided by the INRA Bordeaux, were used as a comparison. These samples possessed known haplotypes, previously described in Dumolin-Lapègue et al. (1997) and Petit et al. (2002a).

### 2.2. Laboratory procedures

After sample collections, plant material was transferred to the laboratory and frozen at -80°C. Subsequently, it was freeze-dried in vacuum and DNA was extracted using the DNeasy 96 extraction kit (Qiagen, Hilden, Germany). Polymerase Chain Reactions (PCR) were carried out for the amplification of ten chloroplast DNA microsatellites (cpSSRs). These were chosen according to a previous phylogenetic study in oaks (Deguilloux et al., 2004). The analyzed loci included ccmp2, ccmp6 and ccmp10, designed on *Nicotiana tabacum* (Weising et al., 1999), and µcd4, µcd5, µdt1, µdt3, µdt4, µkk3 and µkk4, designed on *Q. petraea* and *Q. robur* (Deguilloux et al., 2003). Primers from these loci were fluorescently labelled with blue (FAM; loci ccmp10, µdt1 and µcd4), green (HEX; loci ccmp2, ccmp6, µdt3 and µdt4) or yellow colour (Atto550; µcd5, µkk3 and µkk4) and were mixed in one multiplex. For the PCR reaction, the SuperHot Mastermix (Genaxxon, Biberach, Germany) a premixed mastermix including all PCR components except DNA template and primers was used. Reaction volume was set to 10µl, comprised of 5 µl reaction mastermix (2x), 1 µl of primer mix (0.2 µM of each primer pair in the mastermix), 2 µl water and 1 µl diluted DNA (ca. 4 ng/µl). The PCR-programm included following steps: (1) denaturation at 95°C for 15 min; (2) 30 cycles with a denaturation step at 94°C for 30s, primer annealing at 46°C for 1 min and 30s and an elongation step at 72°C for 30s; (3) Elongation at 72°C for 10min and (4) a final step at 60°C for 30min. Allele scoring was performed by means of a capillary electrophoresis using an ABI Prism 3100 genetic analyzer and the software GeneMapper (Applied Biosystems). Repeats of capillary electrophoresis were carried out by reanalyzing representative subsamples from different 96-well plates together the reference samples of the most common central European haplotypes of white oaks on a single 96-well plate.

### 2.3. Data analysis

Chlorotypes were defined as different combinations of alleles at the 10 analyzed cpSSR loci. For chlorotype designation, loci were sorted alphabetically and alleles were sorted by increasing size. In order to define the phylogenetic relationships among the chlorotypes, a minimum spanning tree based on allele length differences was constructed using the software Arlequin 3.5.1.3 (Excoffier and Lischer, 2010). The software HapStar was then applied to plot the minimum spanning tree (Teacher and Griffiths, 2011).

Genetic variation within and among stands was surveyed by means of various differentiation measures. The software programm PermutCpSSR 2.0 (available at

http://www.pierroton.inra.fr/genetics/labo/Software/PermutCpSSR/index.html) was used to investigate haplotypic diversity within populations ($h_S$) and overall ($h_T$). Pairwise non-parametric Mann-Whitney tests were carried out (using a standard programme available online at http://elegans.som.vcu.edu/~leon/stats/utest.cgi) in order to test whether $h_S$ differences between (1) species, (2) groups of stands belonging to different age classes or (3) groups of stands belonging to different biogeographical subregions are significant. The data sets used for the tests consisted of population specific haplotypic diversities ($h_k$) as described in (Pons and Petit, 1996) which are averaged in order to find out within population diversity of a particular group of populations. Normal approximations were made when sample sizes exceeded 20 populations, following the recommendations of Sokal and Rohlf (1995). Comparisons were made both within and across species. Moreover, using the programm PermutCpSSR 2.0, differentiation among populations ($G_{ST}$) and equivalent coefficient of differentiation ($R_{ST}$), which takes into account mutational differences among haplotypes, were calculated. In order to test for phylogeographic structure, a statistical test was employed following Pons and Petit (1996) in order to compare the values of $G_{ST}$ and $R_{ST}$. According to Lynch and Crease (1990) and Pons and Petit (1996), if the different chlorotypes are randomly distributed, the values of these two measures are equal. On the contrary, if phylogeographic structure occurs, e.g. each chlorotype has a specific geographic distribution, $R_{ST}$ is larger than $G_{ST}$. To test for phylogeographic structure, the measured $R_{ST}$ values were compared with the distribution of values obtained by 1000 random permutations of haplotype identities among populations using the software PermutCpSSR 2.0. Calculations of genetic variation measures, as well as the test for phylogeographic structure were carried out both at the within species level (with populations with at least 3 individuals of the analyzed species) and across species (without considering species identity in populations).

To further investigate interspecific differentiation, an Analysis of Molecular Variance (AMOVA) based on Excoffier et al. (1992) and Excoffier (2003) was conducted using Arlequin 3.5.1.3. The analysis was based on the haplotypic format and on the number of different alleles ($F_{ST}$-like), following the instructions of the program manual (available online at http://popgen.unibe.ch/software/arlequin35/man/Arlequin35.pdf). Species were treated as groups resulting in three hierarchical levels of differentiation: among species (fixation index $F_{CT}$), among populations within species (fixation index $F_{SC}$) and within populations (fixation index $F_{ST}$). Significance tests were carried out for each one of the corresponding fixation indices by applying 10,000 permutations of (1) populations among species for testing significance of $F_{CT}$, (2) haplotypes among populations within species for testing significance of $F_{SC}$ and (3) haplotypes among populations across species for testing significance of $F_{CT}$.

Furthermore, in order to examine the degree of cpDNA sharing between species, introgression ratios (IG) according to (Belahbib et al., 2001) were calculated by examining the study species pairwise. The introgression ratio was calculated as the ratio between the interspecific genetic identity and the mean of the intraspecific genetic identities. The intraspecific genetic identity equals the sum of squares of the frequencies of all haplotypic variants within each species in the whole study area and the interspecific one is defined as the sum of products of each haplotypic variant between species (Dumolin-Lapègue et al., 1999). The introgression ratio between species 1 and 2 was calculated using the following formula: $IG = 2J_{12}/(J_1+J_2)$, where $J_{12}$= interspecific genetic identity and $J_1$, $J_2$= intraspecific genetic identities. An IG score equal to zero indicates total absence of introgression between the two examined demes, whereas an IG score equal to one means that all the variation is shared between them.

Finally, it was aimed to investigate the spatial pattern of haplotypic differentiation. It has been repeatedly shown that the spatial distribution of chloroplast DNA variation in oaks is discontinuous (e.g. Petit et al., 2002a; Grivet et al., 2006; Neophytou et al., 2011). Thus, it was sought to identify differentiated groups and to set significant barriers between highly differentiated groups of populations. For this purpose, a Spatial Analysis of Molecular Variance (SAMOVA, Dupanloup et al., 2002) was employed to define geographically homogenous groups of populations, which show maximum genetic differentiation from each other. To do so, the SAMOVA algorithm was used to maximize the value of differentiation among groups ($F_{CT}$). The predefined number K of such groups was set from 1 to 10. 10,000 iterations for each K were run for each of 100 simulated annealing processes. Subsequently, $F_{CT}$ values among different K configurations were compared, in order to choose the optimal group number (K). On the one hand, the highest $F_{CT}$ represents the optimal configuration. On the other hand, higher K configurations with single population groups cannot provide the group structure (Heuertz et al., 2004). Thus, among all K values without single population groups, the K value resulting in the highest $F_{CT}$ was chosen as the optimal clustering solution. Both SAMOVA and Barrier Analysis were carried out (1) by treating all species as a single gene pool and (2) by examining *Q. robur* and *Q. petraea* separately (the sampling size of *Q. pubescens* was not sufficient in order to carry out a meaningful separate analysis). Finally, a G-test of independence (Sokal and Rohlf, 1995; pp. 738-739) was carried out in order to test the null hypothesis of genetic homogeneity (in terms of SAMOVA group assignment) across biogeographical subregions (definition in Table 1).

## 3. RESULTS

### 3.1. Chlorotype designation and phylogenetic relationships

All analyzed chloroplast microsatellite loci were variable possessing 2, 3 or 4 different alleles each. Different combinations of these allelic variants resulted in 19 chlorotypes in total (Table 2). Among them, four were restricted to refugial areas and were thus given a distinct designation (chlorotypes 21-24), whilst chlorotypes found in the Upper Rhine Valley were designated with numbers 1 to 15. Analysis of DNA samples possessing a known chlorotype based on an analysis of Restriction Fragment Length Polymorphisms (RFLP; data presented in Petit et al., 2002a) led to assignment of them to the cpSSR chlorotypes described in the current study. In particular, RFLP derived chlorotypes 7 and 5, belonging to the 'Balkan' maternal lineage A (according to Petit et al., 2002a) were assigned to the cpSSR based chlorotypes 6 and 8 respectively. Regarding the 'Iberian' lineage B, RFLP based chlorotypes 10 and 11 were both assigned to chlorotype 12 of the present study. A one by one assignment could be made for the RFLP derived chlorotype 12, which corresponded to the chlorotype 13 of this study. Finally, chlorotype RFLP based chlorotype 1 of the 'Italian' lineage C could be matched to chlorotype 9 of the present study (Table 2).

The five aforementioned cpSSR haplotypes (6, 8, 9, 12 and 13) were among the most common in the Upper Rhine Valley. In particular, chlorotype 6 was the most prevalent in the region with a frequency of over 45% in each species (Table 3). This chlorotype was also highly frequent in the four Balkan populations and was absent from the four Iberian populations of this study, thus providing further support of its Balkan origin. The second most frequent chlorotype in the Upper Rhine was chlorotype 9 with a frequency of at least 30% in each species. Among the refugial populations, it was also found in the Italian populations, which is in accordance with its assignment to lineage C. Chlorotype 12 was the third most common in the Upper Rhine, occurring with frequencies of around 10% in *Q. robur* and *Q. petraea*. It was also found in the Iberian populations, supporting its inclusion to the lineage B. The remaining chlorotypes presented lower frequencies (Table 3). Chlorotype frequency data for each stand are given as supporting information (Supplementary data in electronic version; Table S2).

Analysis of the phylogenetic relationships by means of a Minimum Spanning Tree (Fig. 1) allowed separation of the main maternal lineages, as these had been described in Petit et al. (2002a). Chlorotypes 3, 12, 13, 15, as well as the strictly refugial 23 and 24 formed a separate cluster. Given their geographic distribution, they were assigned to the lineage B, with an Iberian refugial origin and a western to central European distribution (Petit et al., 2002a). Chlorotypes 1, 2, 5, 9 and 10 also clustered together. Evidence from the reference samples and from the refugial populations, as well as consideration of the Lineage C geographic distribution (Italy to Scandinavia; Petit et al., 2002a), supports their assignment to this lineage. Similarly, an inclusion of chlorotypes 4, 6, 7, 8, 11 and 14 to lineage A is suggested by the data. Finally, chlorotypes 21 and 22, found exclusively in Balkan and Italian refugial populations, formed a distinct cluster, separated by three mutational steps from other chlorotypes of lineage A. This suggests its inclusion to lineage E, which is widespread from Italy to Russia and is absent from Middle Europe.

**3.2. Haplotypic variation and interspecific introgression**

Computation of diversity measures indicated a lower within-stand ($h_S$) than overall haplotypic variation ($h_T$) for all species (Table 4). This indicates that stands tend to be relatively homogenous, but differ from each other regarding their haplotypic composition. This was also reflected into the high values of differentiation measures. $R_{ST}$ was higher than $G_{ST}$ suggesting phylogeographic structure, but this difference was not significant, as revealed by the permutation test for phylogenetic structure. However, when species were split to stands less than 70 and over 100 years old, phylogenetic structure was significant in the second group for *Q. petraea*. Moreover, probability value of the test was relatively high ($P = 0.935$) when this was performed based on all populations and without taking species into consideration (Table 4). No significant pairwise differences of diversity were found between population groups belonging to different biogeographical subregions, either within or across species. By comparing all 53 *Q. robur* populations with all 31 of *Q. petraea*, we found a significantly higher haplotypic variation in the former species ($z = 0.2317$, $P = 0.010$). Furthermore, significant differences of $h_S$ values were observed between age classes in *Q. robur*. In particular, the young age class (up to 70 years) measuring $h_S = 0.495$ was significantly more diverse from the stands older than 100 years in *Q. robur*, which showed a value of $h_S = 0.203$ ($z = 3.0869$, $P = 0.001$). Given that the differences between the age classes of 100-130 and 130-200 were very limited and non-significant in either species, we pooled these two groups into one group consisting of stands older than 100 years. Differences among age classes were non-significant in *Q. petraea* (U = 140.0, P = 0.211).

A high degree of chlorotype sharing among the three species was highlighted by the Analysis of Molecular Variance (AMOVA). The analysis revealed minimal ($F_{CT}$ was effectively equal to zero) and non-significant differentiation due to partition into species groups (Table 5). On the contrary, subdivision into different populations within species was a highly significant $F_{SC}$ value equal to 0.638, accounting for 63.80 % of the total molecular variance. Variation within populations was also highly significant ($F_{ST}$ = 0.630) and corresponded to 36.20 % of the total molecular variance. The high degree of interspecific chlorotype sharing was also shown by elevated introgression ratios (IG), which almost approached the maximum value of 1.0. In particular, IG value was 0.987 between *Q. robur* and *Q. petraea*, 0.965 between *Q. robur* and *Q. pubescens* and 0.940 between *Q. petraea* and *Q. pubescens*.

By presenting frequency data from each population on a map, a distinctive geographic pattern of the different lineages was observed. This spatial pattern was shared among the species (Fig. 2). In particular, chlorotype 6 (Balkan lineage A) was prevalent in most stands over a large area in the middle of the Upper Rhine Valley. Chlorotype 9 (Italian lineage C) was highly frequent in the northern part of the study area, starting from the outskirts of Karlsruhe and reaching stands close to Mannheim. It was also dominant in the southern part of the Upper Rhine Valley, in areas between Freiburg and Basel (Switzerland). In both cases, an admixture of other chlorotypes (mainly chlorotype 6) could be observed. Chlorotype 12 (Iberian lineage B) was dominant in the north-western part of the study area, at the piedmont of the Vosges. No admixture of other chlorotypes could be observed in the sampled stands. Chlorotypes 8 (lineage A) and 15 (lineage B) presented a more restricted distribution. Chlorotype 8 was confined to two stands in the middle of the study area, while chlorotype 15 was present in four oak stands between Offenburg and Karlsruhe.

### 3.3. Spatial genetic structures

The results of the Spatial Analysis of Molecular Variance (SAMOVA) reflected the aforementioned spatial structure. The optimal number of groups varied from 3 to 5, depending on the analysis configuration. Without considering species identity, application of the SAMOVA algorithm resulted in four groups with very high and significant genetic differentiation ($F_{CT}$ = 0.598, P = 0.001). By carrying out the analysis only in *Q. robur*, the optimal number of groups was 5 ($F_{CT}$ = 0.510, P = 0.001), while in *Q. petraea* it was 3 ($F_{CT}$ =

0.749, P = 0.001). In all cases, three population groups with a consistent haplotypic composition and geographic distribution were revealed. First, one group with high frequency of the Italian chlorotype 9 and discontinuous distribution including the northernmost and the southernmost populations was identified (Group 1 in Fig. 3). A second group corresponded to the central populations where the Balkan chlorotype 6 was prevalent (Group 2 in Fig. 3). A third group included populations with chlorotype 12 in the northwest (Group 3 in Fig. 3). When analyzing all three species without taking their taxonomic identity into account an additional small group (Group 4 in Fig. 3) was observed. This group included two *Q. robur* populations dominated by chlorotype 15 of the Iberian maternal lineage. The same observation was made when *Q. robur* was treated separately. Finally, a fifth group – also consisting of two populations – was observed only in *Q. robur*. This was characterized by the presence of chlorotype 8. Details about frequency distribution of chlorotypes within each one of the derived groups are given in Table 6.

With respect to site conditions, a heterogeneous frequency distribution of SAMOVA groups among different biogeographic subregions was observed (Table 7). In particular, most of the stands assigned to the SAMOVA group 1 occurred on Early Pleistocene plains and dunes. On the contrary, most stands of the SAMOVA group 2 occurred on Late Pleistocene plains and Holocene floodplains. These frequency differences of the two SAMOVA groups among biogeographic subregions was highly significant ($G_{adj}$ = 33.02, P < 0.001). Due to the limited sample size, SAMOVA groups 3 and 4 were not treated by means of a G-test of independence.

## 4. DISCUSSION

### 4.1. Upper Rhine Valley, a crossroads of oak post-glacial migration

Based on chloroplast microsatellite variation, we performed a detailed phylogeographic analysis of the indigenous oak species in the Upper Rhine Valley. The used marker set was adequate to differentiate chlorotypes and maternal lineages occurring in the region. We could assign the most common chlorotypes of our study to those found in previous large scale studies, based on the RFLP technique, and we could resolve the three maternal lineages – A, B and C according to Petit et al. (2002a) – originating from the three glacial refugia of the European continent (Balkan, Iberian and Italian Peninsula). Two additional chlorotypes found in Balkan refugial populations of the present study were well differentiated and may belong to the lineage E, which is restricted to south-eastern Europe and the Italian Peninsula. Along with lineage D, which occurs only in the Iberian Peninsula, this lineage seems not to be involved in post-glacial migration in the Upper Rhine Valley and the western part of Central Europe in general (Petit et al., 2002a). Maternal lineages A, B and C were represented by several chlorotypes in our study area, showing that extant populations originate from all three refugia. The Upper Rhine Valley appears to be a hot-spot of chloroplast DNA diversity for the European white oaks and a crossroads of post-ice-age recolonization pathways.

Contemporary spatial distribution of chlorotypes in the Upper Rhine fits large scale patterns across Europe. In particular, a continuous occurrence of chlorotype 6 of the Balkan lineage A (chlorotype 7 according to Petit et al., 2002a) in the central part of our study between Haguenau (northern Alsace) and Freiburg (southern Baden) supports that this area was a part of a south-westward recolonization pathway of this chlorotype along a zone north of the Alps. Such a migration route is highly supported by large scale data (Petit et al., 2002a). The Vosges, dominated by chlorotype 12 (corresponding to chlorotypes 10 or 11 of the Iberian lineage B, according to Petit et al., 2002a), should have posed an obstacle to a further migration of chlorotype 6 towards northwest. Furthermore, the distribution of chlorotype 9 (chlorotype 1 according to Petit et al., 2002a) of the Italian lineage C in our study area is in concordance with its broader distribution beginning from the Apennine Mountains in the South and reaching the Scandinavian Peninsula in the North (Petit et al., 2002a). Its discontinuous distribution in our area may be due to its delayed arrival in the area. Indeed, pollen analyses support that oaks were able to cross the Alps only during the Atlantic period (7500-5000 years BP) under favourable climatic conditions (Brewer et al. 2002). Thus, oaks with chlorotype 9 could only establish themselves in gaps, which had not been filled during the earlier migration of lineage A.

### 4.2. Convergence of post-glacial migration pathways results in high genetic variation of chloroplast DNA

The complex Holocene migration history of the oaks in the Upper Rhine Valley and particularly the admixture of different maternal lineages might have largely determined chloroplast DNA diversity levels of the Upper Rhine oak forests. A particularly high within population diversity was found in *Q. robur* with $h_S$ reaching the value of 0.346 (measurement based on 53 populations). This is remarkably higher than the value of 0.183 found in *Q. robur* in the pan-European study of Petit et al. (2002a) which included 2600 populations. It is also higher than the respective values found for this species in various studies at the regional scale. For instance, König et al. (2002) measured an $h_S$ value of 0.239 across Central Europe (231 populations), Olalde et al. (2002) measured an $h_S$ value of 0.205 across the Iberian Peninsula (37 populations), while Petit et al. (2002c) found an $h_S$ equal to 0.177 across France (286 populations). Diversity ($h_S$) values for *Q. robur* measured in the British Isles ($h_S$ = 0.224 based on 89 populations; Cottrell et al., 2002), the Northwestern Balkans ($h_S$ = 0.233 based on 88 populations; Bordács et al., 2002) and the Northeastern Balkans ($h_S$ = 0.196 based on 26 populations; Moldovan et al., 2010) were also lower than the value observed in the Upper Rhine Valley in the present study. In *Q. petraea*, genetic variation within stands was also relatively high ($h_S$ = 0.162). Among all previously cited studies, only König *et al.* (2002) reported a slightly higher $h_S$ value ($h_S$ = 0.164 based on 172 populations across Central Europe). All other studies measured values between 0.088 and 0.156. Due to limited sampling and the marginal distribution of *Q. pubescens* in the region, results of the present study about within stand variation of this species are not conclusive.

The intermixing of different chlorotypes due to the meeting of several different migration routes might have also resulted in looser genetic structures. Although a broad scale geographic pattern can be recognized, $G_{ST}$ values were low in comparison to other studies and phylogeographic structure was non-significant, with the exception *Q. petraea* stands older than 100 years. A lack of significant phylogeographic structure in the European white oaks has been reported by other studies as well (Grivet et al., 2006). On the contrary, chloroplast DNA of oaks from regions with Mediterranean climate and at refugial sites in general displays a more pronounced phylogeographic structure. Unlike Central Europe, oaks in the Mediterranean (and other regions with Mediterranean climate like California) persisted in multiple refugia resulting in an *in-situ* differentiation of cpDNA during the ice age and probably throughout the Quartenary. This might have finally resulted in patchy distribution and significant phylogeographic structure with well differentiated lineages occurring relatively close to each other and a remarkably higher total chlorotypic variation (Grivet et al., 2006; López de Heredia et al., 2007; Neophytou et al., 2011). In contrast, the recent recolonization of central Europe by a limited number of chlorotypes, as well as intermixing during this procedure might have contributed to a wide-scale spatial pattern and the lack of phylogeographic structure.

### 4.3. Genetic variation differences among species reflect their distinctive ecological characteristics

Ecological and physiological differences between *Q. robur* and *Q. petraea* probably resulted in the higher genetic variation within stands of the former species. First, compared to *Q.*

*petraea*, *Q. robur* produces larger amounts of acorns and these are preferentially collected and cached by jays (Jones 1959; Bossema, 1979; Ducousso et al., 1993; Petit et al., 2004), thus intensifying seed mediated gene flow at the regional scale. Second, *Q. robur* displays a more pronounced pioneer character forming less dense forests and, in later successional stages, it can be replaced by more competitive tree species including the more shade tolerant *Q. petraea* (Oberdorfer, 1992; Ducousso et al., 1993; Aas, 2008a,b). In contrast, *Q. petraea* forms more closed forest stands which persist for a longer time during succession, preventing external influx of genetic variants. The ecological behaviour of *Q. robur* is likely to have enhanced seed mediated gene flow within species, since *Q. robur* might have been more often involved to colonization of empty or disturbed sites. Along with the present study, results from the literature fully support this hypothesis (Reif & Gärtner 2004). Genetic investigations based on cpDNA haplotypic variation have shown a generally higher within-stand variation in *Q. robur* than in *Q. petraea* across their distribution range. This has been the case whether genetic variation has been measured across Europe (Petit et al., 2002a) or separately at the regional level (König et al., 2002; Olalde et al., 2002; Petit et al., 2002b; Cottrell et al., 2002; Bordács et al., 2002). Moreover, consistently lower differentiation among stands of *Q. robur* has been reported, as a result of more intensive seed mediated gene flow in this species, which is also confirmed by our study ($G_{ST}$ = 0.486 in *Q. robur* vs. 0.769 in *Q. petraea*). The significant phylogeographic structure found in the old *Q. petraea* stands, which are more likely of natural origin, is in agreement with the aforementioned hypothesis.

Despite the diversity differences between *Q. robur* and *Q. petraea*, all three indigenous oak species of the Upper Rhine Valley show a high degree of chloroplast DNA sharing. Besides the high introgression ratios between all species pairs, broad scale genetic structures are common among all three study species. This pattern is frequent among related oak species within sections (e.g. Petit et al., 2002b; Lumaret and Jabbour-Zahab, 2009; Gugger and Cavender-Bares, 2011; Neophytou et al., 2011) and has been attributed to historical hybridization and successive backcrossings. Direct evidence of hybridization events among all three study species has been provided through paternity analyses (Streiff et al., 1999; Curtu et al., 2009; Salvini et al., 2009) and large scale cpDNA variation is shared among white oaks in Europe (König et al., 2002; Olalde et al., 2002; Petit et al., 2002c; Cottrell et al., 2002; Bordács et al., 2002). Thus, we argue that past introgressive hybridization involving all three study species accounts for the observed cpDNA sharing in the Upper Rhine Valley.

### 4.4. Coincidence of site condition differentiation and genetic variation may reflect a stepwise migration, but also early human land use

Interestingly, there is a remarkable difference of site conditions between lowland areas occupied by the ‘Balkan’ chlorotype 6 and those occupied by the ‘Italian’ chlorotype 9 in our study. The former can be mostly found on periodically flooded Late Pleistocene plains and Holocene plains, while the latter is rather distributed on well drained soils of Early Pleistocene plains and dunes. This leads to the hypothesis that these two distinct biogeographical subregions were colonized in two different recolonization waves and possibly by different species, even if subsequent introgression has led to a compensation of chlorotype frequencies among species. It is likely that *Q. robur*, being more pioneer and flood-tolerant, first colonized the floodplains during the Early Holocene originating from the east. It is also possible that *Q. petraea* also emerged during this period and partly replaced *Q. robur* through pollen swamping, leading to introgression of cpDNA haplotypes from *Q. robur* into *Q. petraea*. Later on, probably during the Atlantic period (7500-5000 years BP), characterized by a climate warmer than today, conditions on the well drained Early Pleistocene plains and dunes were favourable for a wide dispersion of *Q. pubescens*. It is possible that the northward recolonization wave, comprising the ‘Italian’ chlorotype 9, was mainly formed by this species. Petit et al. (2002b) also support this hypothesis, based on cpDNA data from Switzerland. Onset of cooler climate during the Subboreal (5000-2500 years BP) should then have led to contraction and introgression of *Q. pubescens* forests by other oak species, probably *Q. petraea*, which frequently hybridizes with *Q. pubescens* (Curtu et al., 2009; Salvini et al. 2009). However, introgression might have also taken place in the opposite direction, since chlorotypes 6 and 9 are almost equally represented in *Q. pubescens*.

Human activities, which have been intensive in our study area for at least 7500 years (Küster, 1996; Lang et al., 2003; Zolitschka et al., 2003; Houben et al., 2006), may have additionally influenced oak genetic variation. Each biogeographical subregion offered different suitability for agricultural use with non-flooded Early Pleistocene plains and dunes being the most suitable for such activities. This has resulted in different histories of deforestation and reforestation since the Neolithic Age in each one of the biogeographic subregions. These fluctuations in land-use may have also influenced the succession of colonization or recolonization of areas by oaks, having an additional contribution to the high haplotypic differentiation among biogeographic subregions.

### 4.5. Intensification of human mediated seed transfer likely accounts for significantly higher genetic variation within young *Q. robur* stands

Furthermore, seed and seedling transfer have possibly resulted in an increase of intrapopulation variation and a decrease of interpopulation differentiation. Such an impact might have been stronger on *Q. robur* stands, in comparison with *Q. petraea*. First, *Q. robur* is spread in the lowland regions where land use, resulting in deforestations and reforestations, has been more intensive over a long time span. Second, it is more likely that *Q. robur* populations are artificially established using allochthonous seed material. König et al. (2002) found that among 181 *Q. robur* stands included in their study in Germany only 42% were classified as autochthonous. The respective percentage in *Q. petraea* was 82%. Even today, plantation is the major means of regeneration in this species, while natural regeneration is more frequent in *Q. petraea*. Finally, seed transfers over long distances throughout Europe since the development of rail network during the 19$^{th}$ century mostly involved *Q. robur*, including the introduction of the superb Slavonian oak (*Q. robur* var. *slavonica*, Gailing et al., 2007). In our region, it seems that medium and long distance seed and seedling transfer has been more intensive after the Second World War, obviously due to the improved transportation abilities. Half of our *Q. robur* study stands (26 stands) were established in the last six decades, while the remaining 27 stands have an age of at least 100 years. The former displayed significantly higher haplotypic diversity, which is supportive of this hypothesis.

### 4.6. Conclusions, recommendations for forest management

The high genetic diversity of the indigenous oak forests of the Upper Rhine Valley described in this study highlights the necessity of their conservation. We have shown that extant forests carry the imprints of migration events that took place up to 10,000 years ago. Since their initial recolonization, their broad scale spatial genetic structure in terms of chloroplast DNA has not changed drastically. This provides evidence that the majority of modern oak forests in the Upper Rhine are autochthonous. During their long presence in the region throughout the Holocene, oaks have undergone several ecological changes and have been able to adapt over a large number of generations. Thus, it is crucial to conserve this genetic diversity, which reflects a complex evolutionary history and suggests that extant forests are well adapted to their sites. Besides conservation, results of the present study may also serve forest management. By defining geographic areas with distinctive chloroplast DNA variation patterns we provide a useful tool for this purpose. A comparison to the broad geographic pattern should provide clues, whether a forest stand is indigenous. Finally, a cost-efficient molecular method (based on cpSSRs instead on RFLPs) has been described, which can be used for controlling the origin and transfer of reproductive material. By documenting the genetic identity of seed producing trees, it is possible to detect if reproductive material is of the same origin as declared by the vendor.

**ACKNOWLEDGMENTS**

The current study has been conducted in the frame of the Interreg-IV project "The regeneration of the oaks in the Upper Rhine lowlands", funded by the European Regional Development Fund (ERDF), the regional government authority of Baden-Württemberg in Freiburg (Regierungspräsidium Freiburg; RPF), the National Office of Forests (Office National des Forêts; ONF) in France and the Regional Directory of Food, Agriculture and Forestry of Alsace (Direction Régionale de l'Alimentation, de l'Agriculture et de la Forêt d'Alsace; DRAAF). We express our gratitude to all the colleagues of the ONF, RPF and the FVA who worked for sample collections and laboratory analyses. Finally, we would like to thank Iñaki San Vicente and Marco Cosimo Simeone for providing the oak samples from Spain and Italy respectively.

**TABLES**

Table 1 – Characteristics of the main biogeographical subregions of the Upper Rhine Valley

| | Late Pleistocene plains and Holocene floodplains | Early Pleistocene plains and dunes (Hardtwaldungen) | Vosges and Black Forest foothills |
|---|---|---|---|
| Prevalent soil types | sand or gravel | sand | various, frequently rocky soils |
| Groundwater level | shallow | deep or absent | absent |
| Flooding events | Frequent, at least until the 19th century | none | none |
| Suitability to human settlement | low | high | medium |

Table 2 – Description of the chlorotypes (chloroplast haplotypes) found and comparison to RFLP based haplotypes according to previous wide scale studies (Petit *et al.* 2002). Chlorotypes were defined as different allelic combinations among the analyzed markers. Chlorotypes from the Upper Rhine were given the designation numbers 1-15 and chlorotypes found only in refugial sites were given the designation numbers 21-24.

| Chlorotype | ccmp2 | ccmp6 | ccmp10 | µcd4 | µcd5 | µdt1 | µdt3 | µdt4 | µkk3 | µkk4 | Correspondent haplotype from Petit *et al.* (2002) |
|---|---|---|---|---|---|---|---|---|---|---|---|
| 1 | 234 | 101 | 110 | 92 | 77 | 79 | 122 | 144 | 102 | 114 | |
| 2 | 234 | 101 | 110 | 93 | 77 | 79 | 122 | 144 | 102 | 114 | |
| 3 | 234 | 101 | 110 | 94 | 78 | 80 | 122 | 142 | 102 | 113 | |
| 4 | 234 | 101 | 111 | 93 | 77 | 79 | 122 | 143 | 102 | 113 | |
| 5 | 234 | 101 | 111 | 93 | 77 | 79 | 122 | 144 | 102 | 114 | |
| 6 | 234 | 101 | 111 | 93 | 77 | 79 | 123 | 143 | 102 | 113 | 7 (lineage A) |
| 7 | 234 | 101 | 111 | 93 | 78 | 79 | 123 | 143 | 102 | 113 | |
| 8 | 234 | 101 | 111 | 94 | 77 | 78 | 123 | 143 | 102 | 113 | 5 (lineage A) |
| 9 | 234 | 101 | 111 | 94 | 77 | 79 | 122 | 144 | 102 | 114 | 1 (lineage C) |
| 10 | 234 | 101 | 111 | 94 | 78 | 79 | 122 | 144 | 102 | 114 | |
| 11 | 234 | 101 | 111 | 94 | 78 | 79 | 123 | 143 | 102 | 113 | |
| 12 | 235 | 101 | 110 | 94 | 78 | 80 | 122 | 142 | 102 | 113 | 10/11 (lineage B) |
| 13 | 235 | 101 | 110 | 94 | 78 | 81 | 122 | 142 | 102 | 113 | 12 (lineage B) |
| 14 | 235 | 101 | 111 | 93 | 77 | 79 | 123 | 143 | 102 | 113 | |
| 15 | 235 | 102 | 110 | 94 | 78 | 80 | 122 | 142 | 102 | 113 | |
| 21 | 234 | 101 | 111 | 92 | 77 | 80 | 121 | 142 | 102 | 113 | |
| 22 | 234 | 101 | 111 | 94 | 77 | 80 | 121 | 142 | 103 | 113 | |
| 23 | 235 | 101 | 110 | 94 | 79 | 81 | 122 | 142 | 102 | 113 | |
| 24 | 235 | 101 | 110 | 95 | 78 | 80 | 122 | 142 | 102 | 113 | |

Table 3 – Relative chlorotype frequencies in the Upper Rhine (pro species) and in refugial populations (pooled white oak species). $N_{ind}$ = number of individuals, $N_{pop}$ = number of populations.

| | Upper Rhine populations | | | Refugial populations | | |
|---|---|---|---|---|---|---|
| | *Quercus robur* | *Quercus petraea* | *Quercus pubescens* | Iberian populations | Italian populations | Balkan populations |
| Chlorotype | *$N_{ind}$ = 314, $N_{pop}$ = 63* | *$N_{ind}$ = 214, $N_{pop}$ = 39* | *$N_{ind}$ = 90, $N_{pop}$ = 6* | *$N_{ind}$ = 32, $N_{pop}$ = 4* | *$N_{ind}$ = 15, $N_{pop}$ = 3* | *$N_{ind}$ = 24, $N_{pop}$ = 4* |
| 1 | 0,64% | 0,93% | - | - | 26,67% | - |
| 2 | - | 0,47% | - | - | 26,67% | - |
| 3 | - | 0,47% | - | - | - | - |
| 4 | 0,32% | 0,47% | - | - | - | - |
| 5 | 0,32% | - | - | - | - | - |
| 6 | 45,86% | 46,73% | 51,11% | - | 6,67% | 54,17% |
| 7 | 0,32% | 1,87% | - | - | - | - |
| 8 | 1,27% | - | - | - | - | - |
| 9 | 33,76% | 30,37% | 46,67% | - | 26,67% | - |
| 10 | 0,32% | 1,87% | - | - | - | - |
| 11 | - | 0,47% | - | - | - | - |
| 12 | 8,92% | 14,95% | 1,11% | 3,13% | - | - |
| 13 | 4,14% | 0,47% | 1,11% | 78,13% | - | - |
| 14 | - | 0,93% | - | - | - | - |
| 15 | 4,14% | - | - | - | - | - |
| 21 | - | - | - | - | 6,67% | 20,83% |
| 22 | - | - | - | - | - | 25,00% |
| 23 | - | - | - | 3,13% | - | - |
| 24 | - | - | - | 15,63% | 6,67% | - |

Table 4 – Haplotypic diversity measures within and across species. $N_{pop}$ = number of populations used for the analysis (containing at least three individuals of either species), $h_S$ = haplotypic diversity within populations, $h_T$ = overall haplotypic diversity, $G_{ST}$ = differentiation among populations, $R_{ST}$ = equivalent coefficient of differentiation, $P(R_{ST}>G_{ST})$ = Probability value of the test for phylogeographic structure. Two different age classes, less than 70 and more than 100 years, were treated separately.

| | | $N_{pop}$ | $h_S$ | $h_T$ | $G_{ST}$ | $R_{ST}$ | $P(R_{ST}>G_{ST})$ |
|---|---|---|---|---|---|---|---|
| *Quercus robur* | < 70 years | 26 | 0.495 (± 0.064) | 0.729 (± 0.034) | 0.321 (± 0.077) | 0.391 (± 0.096) | P = 0.448 |
| | > 100 years | 27 | 0.203 (± 0.053) | 0.600 (± 0.058) | 0.662 (± 0.091) | 0.683 (± 0.098) | P = 0.779 |
| | overall | 53 | 0.346 (± 0.046) | 0.673 (± 0.031) | 0.486 (± 0.063) | 0.521 (± 0.069) | P = 0.750 |
| *Quercus petraea* | < 70 years | 14 | 0.218 (± 0.074) | 0.728 (± 0.020) | 0.701 (± 0.102) | 0.789 (± 0.088) | P = 0.932 |
| | > 100 years | 17 | 0.125 (± 0.056) | 0.670 (± 0.075) | 0.814 (± 0.086) | 0.914 (± 0.057) | P = 0.993 |
| | overall | 31 | 0.162 (± 0.045) | 0.702 (± 0.027) | 0.769 (± 0.063) | 0.778 (± 0.070) | P = 0.578 |
| *Quercus pubescens* | | 6 | 0.078 (± 0.056) | 0.596 (± 0.021) | 0.869 (± 0.098) | 0.847 (± 0.101) | P = 0.161 |
| Pooled species | | 90 | 0.290 (± 0.032) | 0.677 (± 0.024) | 0.572 (± 0.046) | 0.619 (± 0.047) | P = 0.935 |

Table 5 – Analysis of Molecular Variance (AMOVA); d.f = degrees of freedom. Negative percentages of variation and fixation indices were turned to zero.

| Source of variation | d.f. | Fixation indices | Percentage of variation |
|---|---|---|---|
| Among species | 2 | $F_{CT} = 0.000$ ($P = 0.797$) | 0.00% |
| Among populations within species | 87 | $F_{SC} = 0.638$ ($P < 0.001$) | 63.80% |
| Within populations | 501 | $F_{ST} = 0.630$ ($P < 0.001$) | 36.20% |

Table 6 – Chlorotype frequencies within the groups revealed by the Spatial Analysis of Molecular Variance (SAMOVA). Results are presented for the analyses across all species and within each of *Quercus robur* and *Quercus petraea*. The spatial distribution of the groups is shown in Fig. 3

| | | Chlorotype | | | | | |
|---|---|---|---|---|---|---|---|
| | Group | 6 | 8 | 9 | 12 | 15 | Other |
| Across species | 1 | 15% | 0% | 77% | 3% | 0% | 4% |
| | 2 | 80% | 2% | 3% | 8% | 1% | 7% |
| | 3 | 0% | 0% | 0% | 100% | 0% | 0% |
| | 4 | 0% | 0% | 0% | 8% | 67% | 25% |
| *Quercus robur* | 1 | 10% | 0% | 84% | 3% | 0% | 2% |
| | 2 | 71% | 1% | 11% | 9% | 2% | 7% |
| | 3 | 0% | 0% | 80% | 10% | 10% | 0% |
| | 4 | 0% | 0% | 0% | 8% | 67% | 25% |
| | 5 | 0% | 50% | 0% | 50% | 0% | 0% |
| *Quercus petraea* | 1 | 2% | 0% | 90% | 0% | 0% | 8% |
| | 2 | 83% | 0% | 3% | 4% | 0% | 10% |
| | 3 | 4% | 0% | 0% | 96% | 0% | 0% |

Table 7 – Frequency of genetically homogenous SAMOVA groups (across species) within each one of the biogeographical subregions (number of stands). See Table 1 for the definition of biogeographical subregions and Fig. 3 for the spatial distribution of the SAMOVA groups.

| | SAMOVA groups (across species) | | | |
|---|---|---|---|---|
| Biogeographic subregion | 1 | 2 | 3 | 4 |
| Late Pleistocene plains and Holocene floodplains | 5 | 27 | 0 | 0 |
| Early Pleistocene plains and dunes (Hardtwaldungen) | 22 | 3 | 0 | 2 |
| Vosges and Black Forest foothills | 8 | 13 | 6 | 0 |

## FIGURES

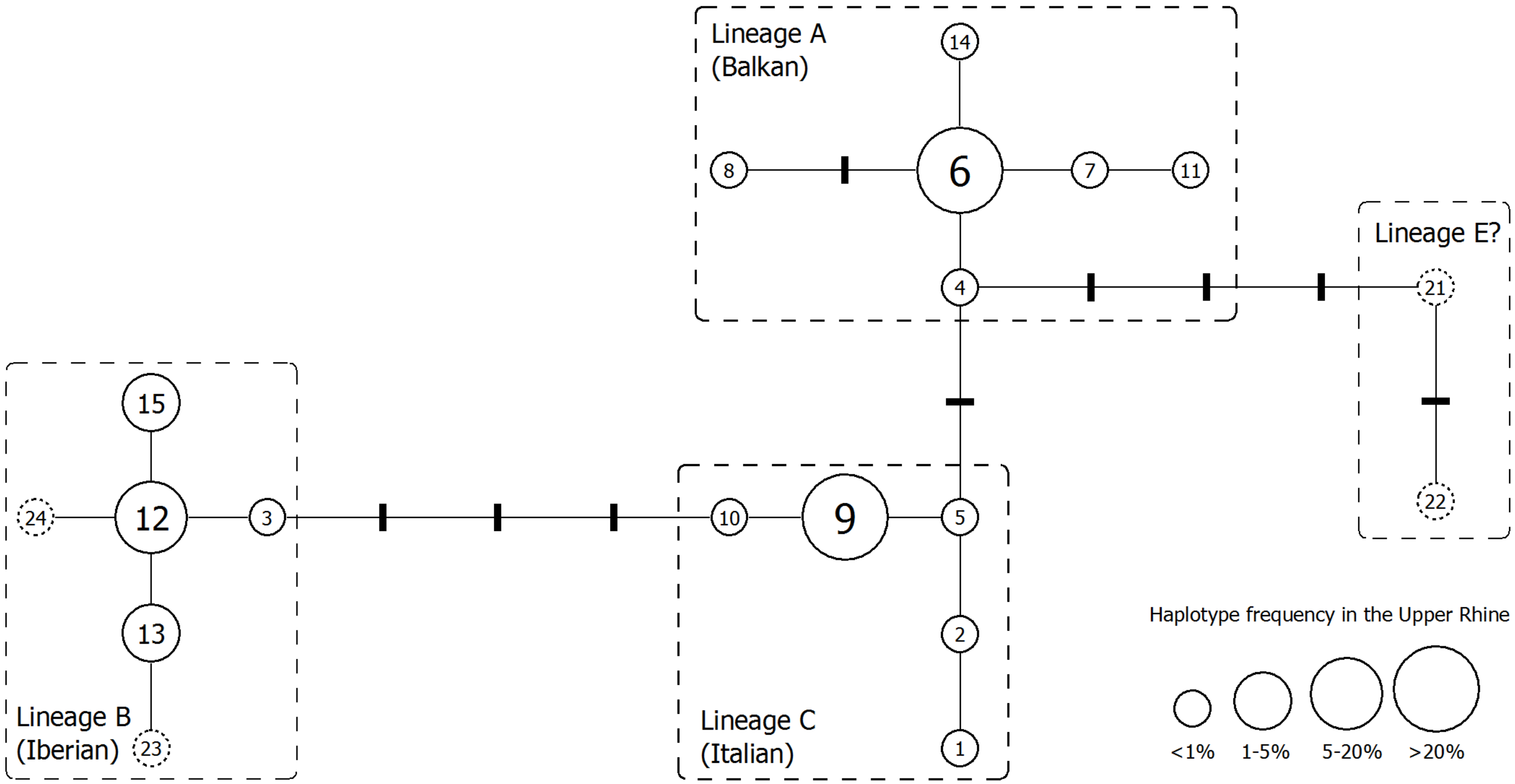

Figure 1 – Minimum Spanning tree of chlorotypes. Chlorotypes are marked with white filled circles. The size of the circles is according to their frequency in the Upper Rhine. Dashed circles indicate chlorotypes found only in refugial populations (21-24). Each segment represents a mutation step (a length difference of one nucleotide). Vertical bars represent missing haplotypes. They were assigned after comparison of the study chlorotypes with reference samples with known chlorotypes.

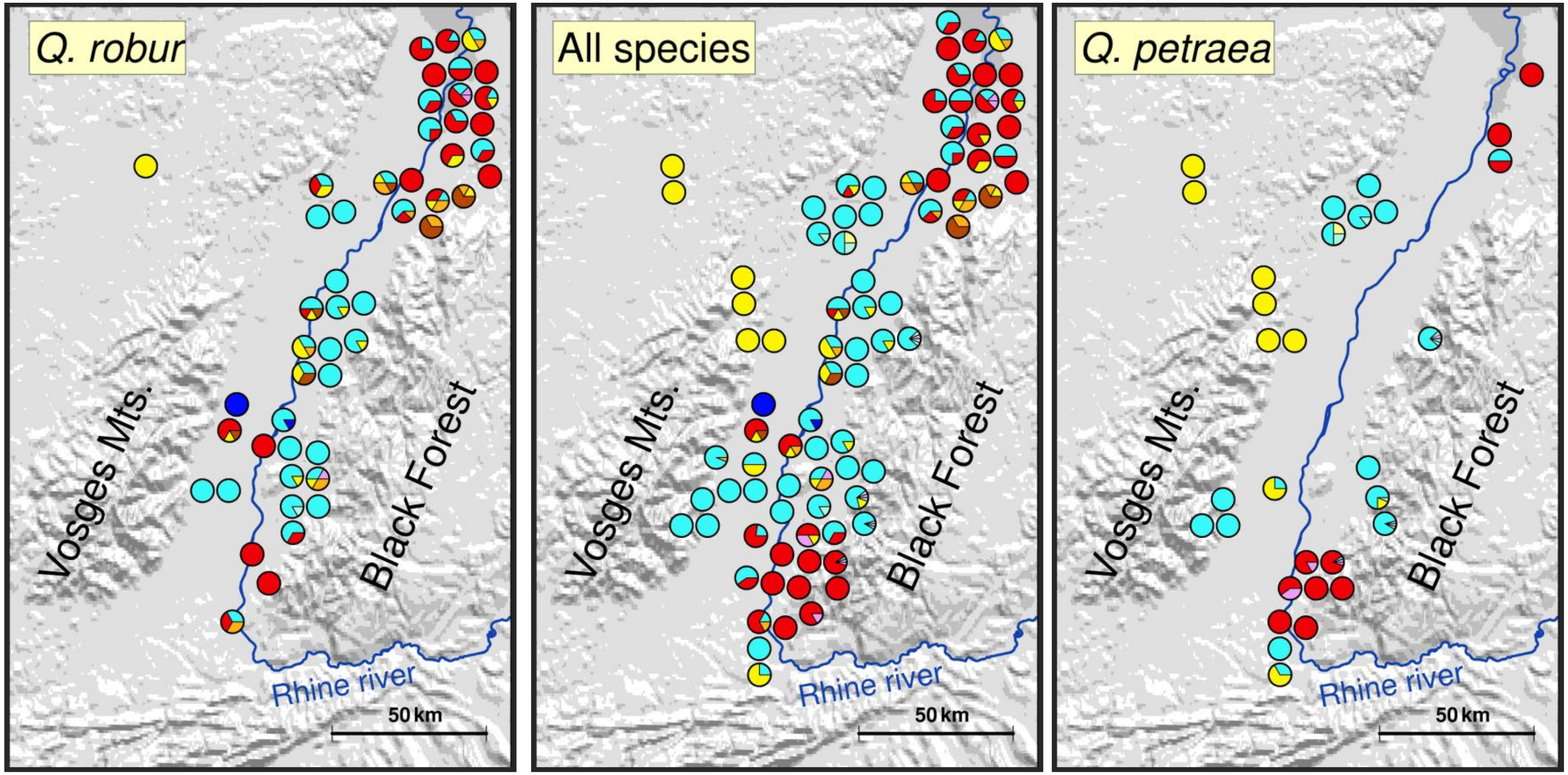

Figure 2 – Spatial distribution of chlorotypes in all populations, irrespective of species identity (to the left) and within each species. Chlorotypes are designated as follows: chlorotype 6 (lineage A): sky blue, chlorotype 8 (lineage A): dark blue, chlorotype 9 (lineage C): red, chlorotype 12 (lineage B): yellow, chlorotype 13 (lineage B): orange, chlorotype 15 (lineage B): brown. For chlorotype description, see Table 2.

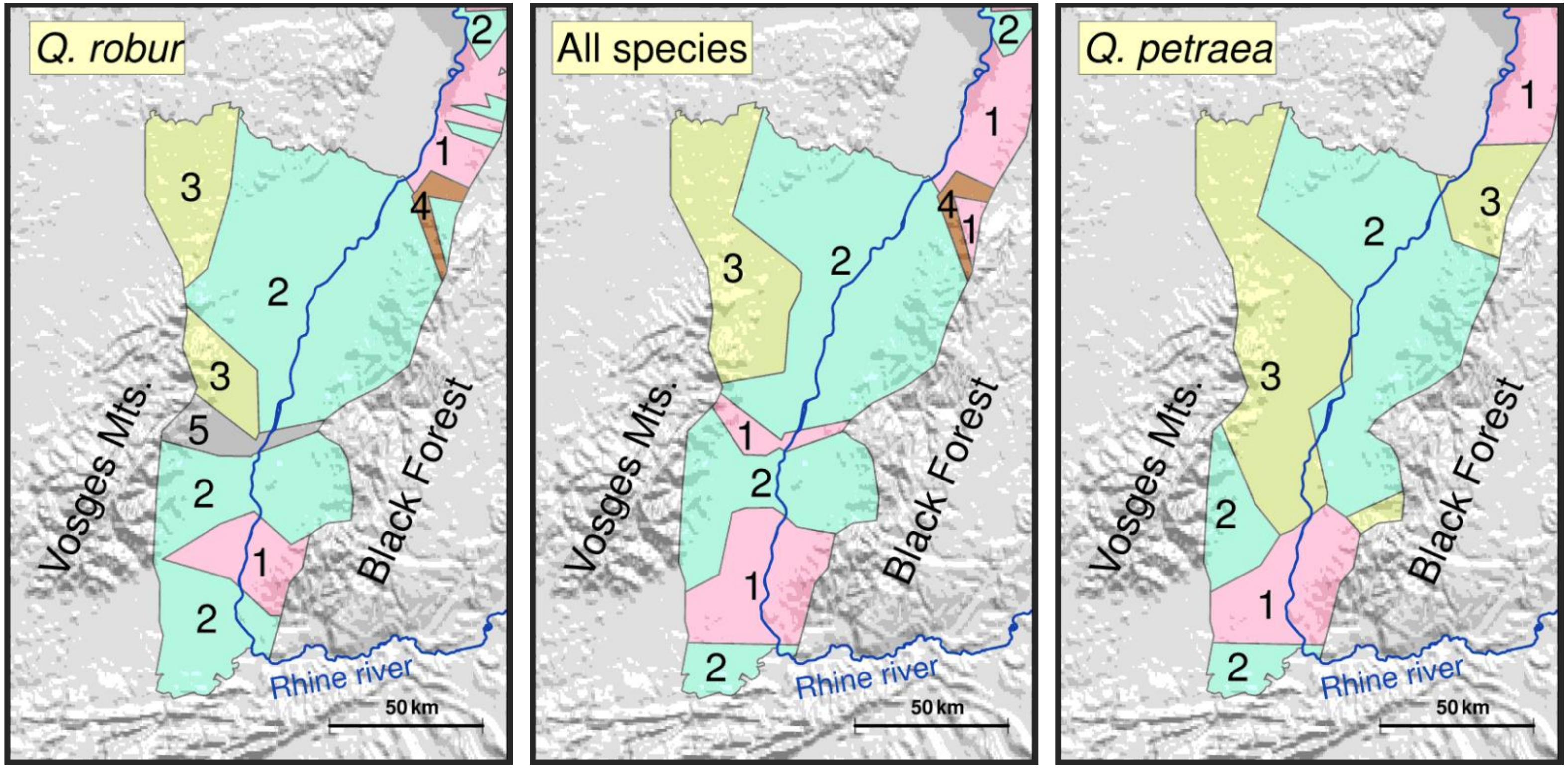

Figure 3 – Spatial genetic subdivision of oak populations according to a Spatial Analysis of Molecular Variance (SAMOVA). Results based on all populations without considering species identity are presented in the left map; results for *Quercus robur* are presented in the middle and for *Quercus petraea* to the right. Three main groups with similar haplotypic constitution and spatial distribution could be observed in all species: a group with high frequency of the Italian chlorotype 9 to the north and to the south (Nr. 1; red), a group dominated by the Balkan chlorotype 6 in the middle and in the southernmost part (Nr. 2; blue) and a group with prevalence of chlorotype 12 of the Iberian maternal lineage (Nr. 3; yellow). A separate group (Nr. 5; violet) including two populations characterized by chlorotype 8 (Balkan lineage) was formed only in *Q. robur*. Group Nr. 4 (brown) dominated by chlorotype 15 (Iberian lineage) included two *Q. robur* populations and thus was not observed in *Q. petraea*.

**SUPPORTING INFORMATION**

Additional supporting information may be found in the online version of this article:

**Table S1** Geographic, taxonomic and forest inventory data of the study stands

**Table S2** Chlorotype frequencies in the study stands